\documentclass[11pt,twoside]{article}

\usepackage{asp2006}
\usepackage{epsf}
\usepackage{psfig}
\usepackage{lscape}

\begin{document}
\title{Galaxy Evolution and Environment}   %%% Fill in title
\author{Pieter van Dokkum \& Ryan Quadri}   %%% Fill in author names
\affil{Yale University}    %%% Fill in author affiliations

\begin{abstract} %%% Abstract to run on from here.
The properties of galaxies are strongly correlated with their
environment,
with red galaxies dominating galaxy clusters and blue galaxies
dominating the general field. However, not all field galaxies
are young: studies of the colors, line strengths, and $M/L$
ratios of massive early-type galaxies at $0<z<1.3$ show that the
most massive galaxies do not seem to care
about their surroundings, and have very similar ages irrespective
of their environment. There is good evidence that the {\em growth}
of these galaxies does continue longer in the field than in clusters,
via (nearly) dissipationless mergers of already old galaxies.
These results are consistent with predictions
of recent galaxy formation models,
which incorporate AGN feedback to suppress star formation in the
most massive halos.
Systematic studies of the relation of galaxies with their
environment beyond $z=1$ are difficult, and still somewhat
contradictory. Intriguingly both the DEEP-2 and VVDS surveys
find that the
color-density relation disappears at
$z\sim 1.3$, unfortunately
just at the point where both surveys become highly
incomplete. On the other hand, clustering studies at $z\sim 2.5$
have shown that red galaxies cluster more strongly than blue
galaxies, implying that the color-density relation was already
in place at that redshift. 

\end{abstract}

%%% MAIN BODY OF TEXT GOES HERE. CONSULT "INSTRUCTIONS FOR AUTHORS USING
%%% LATEX2E MARKUP", SECTIONS 2.3-2.6 FOR HELP WITH EQUATIONS, FIGURES,
%%% AND TABLES.

%\section{}   %%% Top level section head (remove "%" symbol)
%\subsection{}   %%% Second level section head (remove "%" symbol)
%\subsubsection{}   %%% Lowest level section head (remove "%" symbol)
%\section*{}    %%% Unnumbered top level section head (remove "%" symbol)
%\subsection*{}   %%% Unnumbered second level section head (remove "%" symbol)

\section{Introduction}

It has been known for many decades that the properties of galaxies
correlate strongly with their environment. Early studies demonstrated
that galaxies in rich clusters are usually red and often have
early-type morphologies, whereas most galaxies
in the general field are blue spiral galaxies. Dressler (1980)
showed that the morphology of galaxies is a smoothly varying function
of local projected density, setting the stage for many studies over
the following decades which quantified this morphology-density
relation better and tried to identify the cause, or causes.
Weinmann et al.\ (2006) present what could perhaps be described
as the 21$^{\rm st}$ century version of the morphology-density
relation, by quantifying the dependence of galaxy color,
star formation rate and morphology on halo mass, using a large
sample drawn from the SDSS.

A key question is why there are red galaxies at all. Models suggest
that gas should continuously accrete onto halos, shock heat to
the virial temperature, and subsequently
cool, leading to sustained star formation in a disk
and blue colors associated with newly formed stars (e.g.,
Fall \& Efstathiou 1980). The existence of large numbers of
apparently ancient galaxies, in particular in rich clusters,
is not easily explained in this context.
As observations and models have become
more and more sophisticated, this 20+ year old problem has not
gone away, but instead is the topic of extensive debate.

A partial solution, implemented in semi-analytical models of the
1990s, is to postulate that gas cooling is disrupted (due
to ram pressure stripping, harrassment, or other processes)
if a galaxy
is subsumed in a more massive halo and becomes a satellite
(e.g., Kauffmann et al.\ 1999). This ``nurture'' solution
naturally produces red galaxies in clusters (with the exception
of the
central galaxy) and blue field galaxies.
More recently, other mechanisms were added.
In current models star formation is halted when galaxies
exceed a critical mass scale (``nature''),
even if they are the central galaxy in their
halo. This suppression can be due to AGN feedback
(e.g., Croton et al.\ 2006), and/or may occur naturally as a result
of virial shocks (Birnboim, Dekel, \& Neistein 2007).
The observational challenge is to assess the importance of
these various proposed nurture and nature mechanisms as a
function of mass, environment, and cosmic time.

\begin{figure}[h]
\plotone{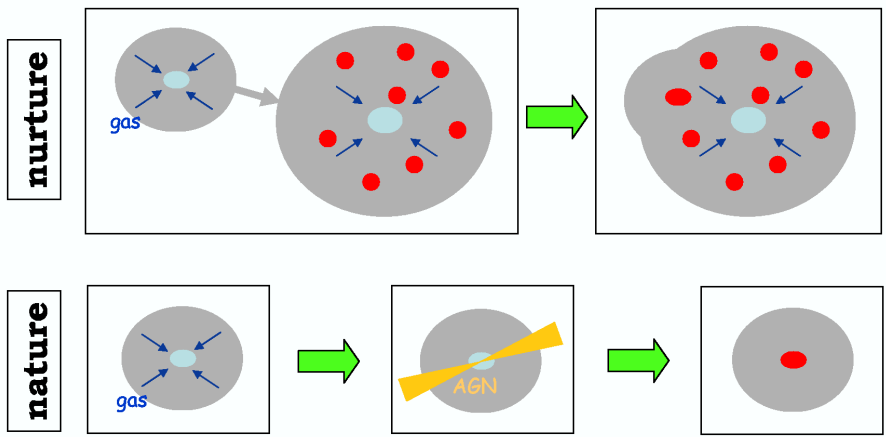}
\caption{The origin of old, red galaxies can be phrased
in ``nurture'' versus ``nature'' terms. Here, ``nurture'' is a
catch-all phrase for the processes which lead to the cessation of
star formation in a galaxy which becomes a satellite in a more
massive halo. ``Nature'' refers to the various processes that
have been proposed to suppress star formation in galaxies when
they exceed a critical halo mass, such as AGN feedback and shock
heating.}
\end{figure}

\section{The Stellar Age of Massive Galaxies}

A long-standing prediction of hierarchical models is that massive
galaxies in low mass halos have very different formation histories
than massive galaxies in high mass halos. Cluster galaxies form
early and their evolution is effectively frozen
shortly after they become part
of a dynamically relaxed massive halo: their velocities are
too high to permit mergers, and various processes rob them of
their ability to form new stars. By contrast, massive galaxies
in the field are expected to grow substantially at late times
through mergers and possibly star formation
(e.g., Kauffmann 1996, de Lucia et al.\ 2006).

We first turn to the star formation epoch of massive
galaxies. The
Kauffmann (1996) models, which did not include AGN feedback,
predicted that the stars in field early-type galaxies\footnote{In
the local Universe, the most massive galaxies
are early-type galaxies -- although this may not hold at higher
redshift.} should
be younger than those in cluster early-type galaxies by
$\sim 4$ Gyr. Models that include AGN feedback (e.g.,
de Lucia et al.\ 2006) predict a much smaller age difference,
of order $\sim 0.7$ Gyr.

Observational studies of early-type galaxies have provided
conflicting evidence on the age difference between massive field-
and cluster early-type galaxies.
Studies of the local Mg$_2 - \sigma$ relation and Fundamental
Plane (FP) find very small
age differences between field and cluster galaxies (e.g., Bernardi et
al.\ 2006). However, Thomas et al.\ (2005) and
Clemens et al.\ (2006) find age differences of 1.5 -- 2 Gyr
from fitting absorption line strengths of nearby galaxies with
complex models that include age, metallicity, $\alpha$-enhancement,
and (in the case of Clemens et al.) carbon enhancement as free
parameters. Studies at higher redshift are also ambiguous:
Treu et al.\ (2005), van der Wel et al.\ (2005), and Rusin \&
Kochanek (2005) find very small age differences from measurements
of $M/L$ ratios, but di Serego Alighieri et al.\ (2006) find
age differences of 3.5 -- 4 Gyr from very similar data.

\begin{figure}[t]
\plotone{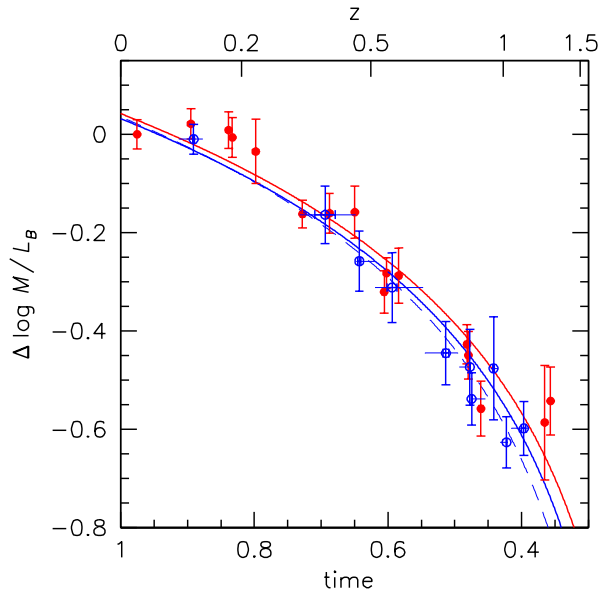}
\caption{Evolution of the $M/L_B$ ratio of early-type
galaxies with stellar masses $>10^{11}\,M_{\odot}$, in the
field (blue) and in clusters (red), from van Dokkum
\& van der Marel (2007). The age difference
between field and cluster galaxies
implied by their luminosity evolution is small at $\sim 4$\,\%.}
\end{figure}

In order to resolve these conflicting results, and to provide
a robust measurement of the stellar age of massive galaxies,
we have recently
compiled FP measurements of cluster- and field galaxies
at $0<z<1.3$ from
the literature, and placed them on a consistent photometric
and dynamical system (van Dokkum \& van der Marel 2007).
The results are shown in Fig.\ 2: the $M/L$ ratios
of massive field and cluster galaxies
evolve in very similar fashion, with little or
no systematic offset. Their age difference is $0.4 \pm 0.2$
Gyr, inconsistent with nurture-only models
and in good agreement with the de Lucia et al.\ (2006)
AGN feedback model.

\section{Dry Mergers}

The ability of current semi-analytical models to reproduce
the stellar ages and colors of massive elliptical galaxies at
$0<z<1$ is a major achievement, but it is somewhat tangential
to one of their central tenets: the prediction that these objects
were assembled relatively recently through mergers.
In clusters mergers should be rare, as the relative velocities
of galaxies are too high. However, in the general field
mergers should continue to the present day. Specifically,
the de Lucia et al.\ (2006) model predicts that massive
ellipticals typically assemble 50\,\% of their final mass
at $z<0.8$, and 20\,\% at $z<0.4$. It is fairly clear,
both observationally and theoretically, that these recent
mergers must have been mostly ``dry''. The star formation
induced by gas-rich mergers
would dramatically change the colors and $M/L$ ratios of
massive elliptical galaxies, inconsistent with observations.
Similarly, semi-analytical models predict that the
most recent mergers of massive galaxies were mostly
gas-poor (e.g., Khochfar \& Burkert 2003; Kang, van den Bosch,
\& Pasquali 2007).

Observationally, there is good evidence that dry mergers
frequently occur in $z\sim 1$ clusters, before they are fully
virialized (van Dokkum et al.\ 1999, Tran et al.\ 2005,
Mei et al.\ 2006). However, 
the importance of dry merging in the general field
is still somewhat
of an open question. Summing up current thinking:
(1) there is no doubt that dry mergers occur; (2)
they likely affected a large fraction of the elliptical galaxy
population; but (3) they may have had only a limited effect on the
evolution of the luminosity function of red galaxies.

Starting with (1),
the recognition of dry mergers goes back to the Hubble atlas and
Arp's catalog (see, e.g., Arp 169 -- 172) --- although back then mergers
were not classified by their aridity. Combes et al.\ (1995)
studied the stellar dynamics of E-E pairs and illustrated some
of the features that occur in mergers of hot stellar systems.
An example of a dry merger at $z=0.1$ is shown in Fig.\ 3:
the ground-based image shows the tidal features, the ACS image
shows the early-type morphologies of the interacting pair, and
the spectrum demonstrates that the red colors stem from old
stars, not from a dust-enshrouded star burst.

\begin{figure}[t]
\plotone{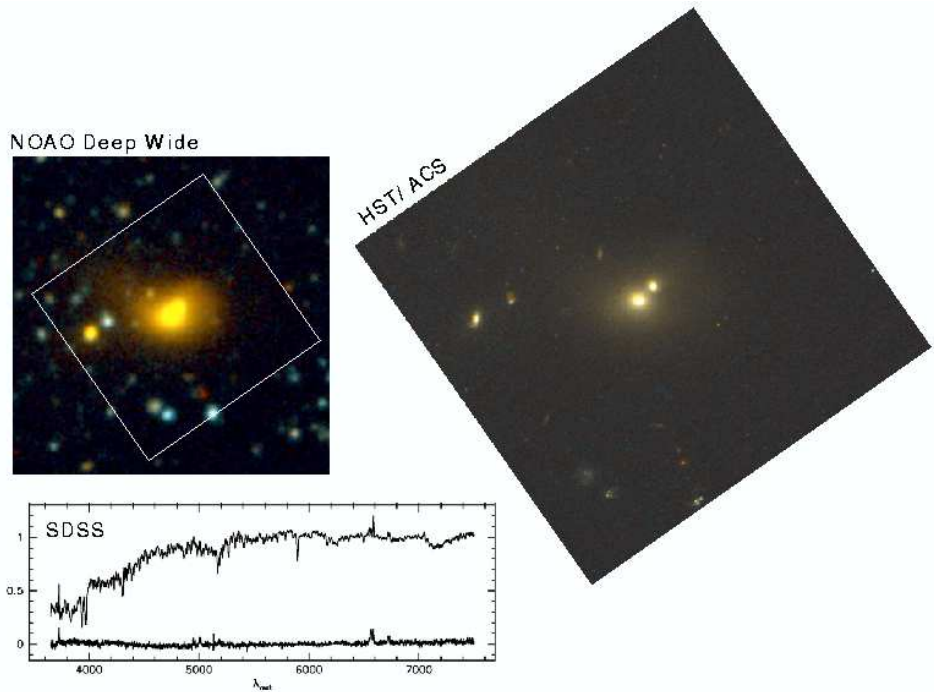}
\caption{Example of a dry merger in the NOAO Deep Wide-Field
Survey, from van Dokkum (2005).
The deep ground-based data were used to identify the merging
pair by its tidal features (which have a surface brightness
of $R \approx 27$). The spectrum shows that the galaxy
light is dominated by old stars (note the LINER features), and
the ACS image shows that, in this case,
the merging galaxies have early-type morphologies.
}
\end{figure}

The difficulty in assessing (2), i.e., how many elliptical galaxies
experienced dry mergers, is that
they progress very rapidly and leave little evidence. Tidal tails
disperse quickly due to the high velocity dispersions of the
progenitors, and the blue, high surface brightness star forming
regions that often grace gas-rich mergers are absent. Detecting
tidal features requires very deep imaging over large
areas. When those depths are reached, red, smooth
tidal features turn out to be very common: in a
sample of 86 bulge-dominated red galaxies in
the NDWF and MUSYC surveys
71\,\% show evidence for tidal interactions (van Dokkum 2005;
see also Schweizer \& Seitzer 1992). In one-third of the tidally
distorted galaxies the merger is still in progress, and the
properties of the merging galaxies could be measured. The
median luminosity ratio of these tidally interacting pairs
is about $1:3$, and the median
color difference is only $-0.02$ in $B-R$. Taken together,
these results indicate that the majority of elliptical galaxies
experienced a merger in its past, that the progenitors were
already mostly ``red and dead'', and that these were not
minor accretion events.

Determining (3), the dry merger rate, is perhaps the
most important question, but it also is the most uncertain.
From the 19 merging pairs in van Dokkum (2005) it was estimated
that dry mergers lead to a mass-accretion rate for galaxies on
the red sequence of $0.09 \pm 0.04$\,Gyr$^{-1}$, that is,
a doubling in mass every 8 Gyr. The uncertainty on this number
is large, and mainly driven by the timescale that needs to be
assumed in the calculation. Bell et al.\ (2006) also use
morphological criteria to identify dry mergers (in the
HST/ACS GEMS field), and estimate that early-type galaxies
have undergone between 0.5 and 2 major dry mergers since
$z\sim 0.7$. Red pair statistics will provide additional constraints
on the dry merger rate. The first study was done by Masjedi
et al.\ (2006a), who determined that the merger rate of
Luminous Red Galaxies (LRGs) in SDSS is very low. However,
this first analysis only considered mergers of LRGs {\em with
each other}, and as these objects  have $L\sim 4 L_*$
they are typically the central galaxies of groups.
Including less luminous neighbors, Masjedi et al.\ (2006b)
derive an accretion rate of $\sim 0.025$\,Gyr$^{-1}$, in better
agreement with van Dokkum (2005), Bell et al.\ (2006),
and theoretical models. We note that
accretion rates of $2.5 - 10$\,\% per Gyr do not imply
a large effect on the luminosity function of red galaxies.
Therefore, recent claims that the high-mass end of the
mass function of red galaxies shows only modest evolution
with redshift (Brown et al.\ 2007; Scarlata et al.\ 2007)
may be entirely consistent with the quoted
studies.

\section{The Color-Density Relation at $z\sim 1$}

The color-density relation might be expected to break down
at some early epoch. Studies that access the star formation epoch
of cluster galaxies may find that red galaxies are rare in dense
environments, and see a rather flat (or possibly even inverted)
color-density relation. The technical challenges are formidable,
as large volumes need to be probed with good redshift
information and a careful handling of systematics.

\begin{figure}[h]
\plotone{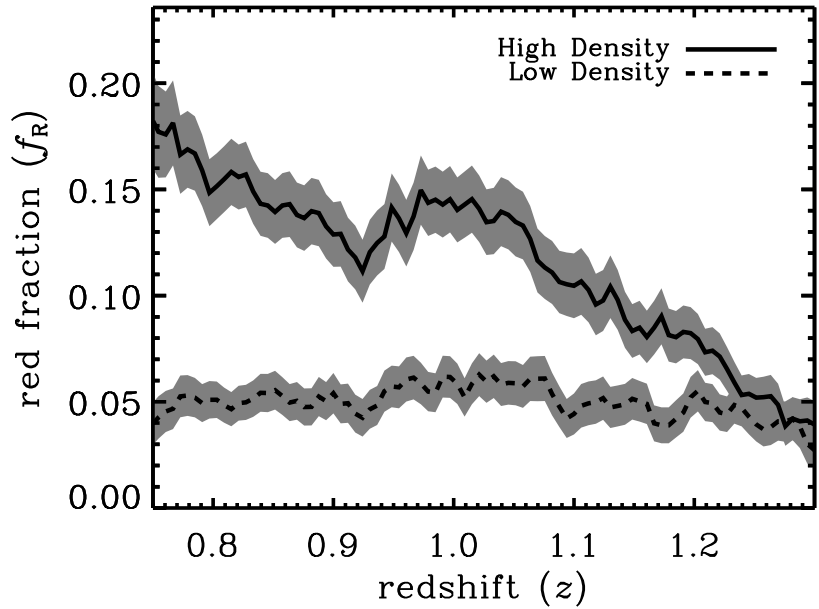}
\caption{The evolution of the contribution of red galaxies to
the total galaxy population, for two different density bins
(from Cooper et al.\ 2007). The red galaxy fraction decreases
strongly with redshift in high density regions,
suggesting that the color-density relation disappears
by $z\sim 1.3$.
}
\end{figure}

Two groups
have recently presented measurements of the color-density relation
out to $z\sim 1.3$, based on very large spectroscopic surveys.
Cucciati et al.\ (2006) use an $I$-selected sample
of 6582 galaxies from
the VIMOS-VLT Deep Survey (VVDS) to determine the evolution
of the color-density relation. They find that the red fraction
increases with density at redshifts up to $z\sim 0.9$, but
then turns over so that it is independent of density
at $z\sim 1.3$. Similarly, Cooper et al.\ find
that the red galaxy fraction is about the same in low-
and high-density regions at $z\sim 1.3$, from a sample of 19464
$R-$selected galaxies obtained in the context of the Keck DEEP2
survey (see Fig.\ 4).
This ``turnover'' redshift seems rather low, in particular
in light of the existence of clusters with well-defined red sequences
at $z\sim 1.3$. Both studies applied
careful corrections for selection effects (a necessity,
as both surveys are highly incomplete at $z>1$), and the fact
that they are in agreement with each other obviously bolsters the
confidence in their findings. 
Further refinement of these intriguing
results may have to await surveys which probe
the redshift range $1<z<2$ with the same robustness as DEEP2
and VVDS have surveyed the Universe out to $z\sim 1$.

\begin{figure}[t]
\plottwo{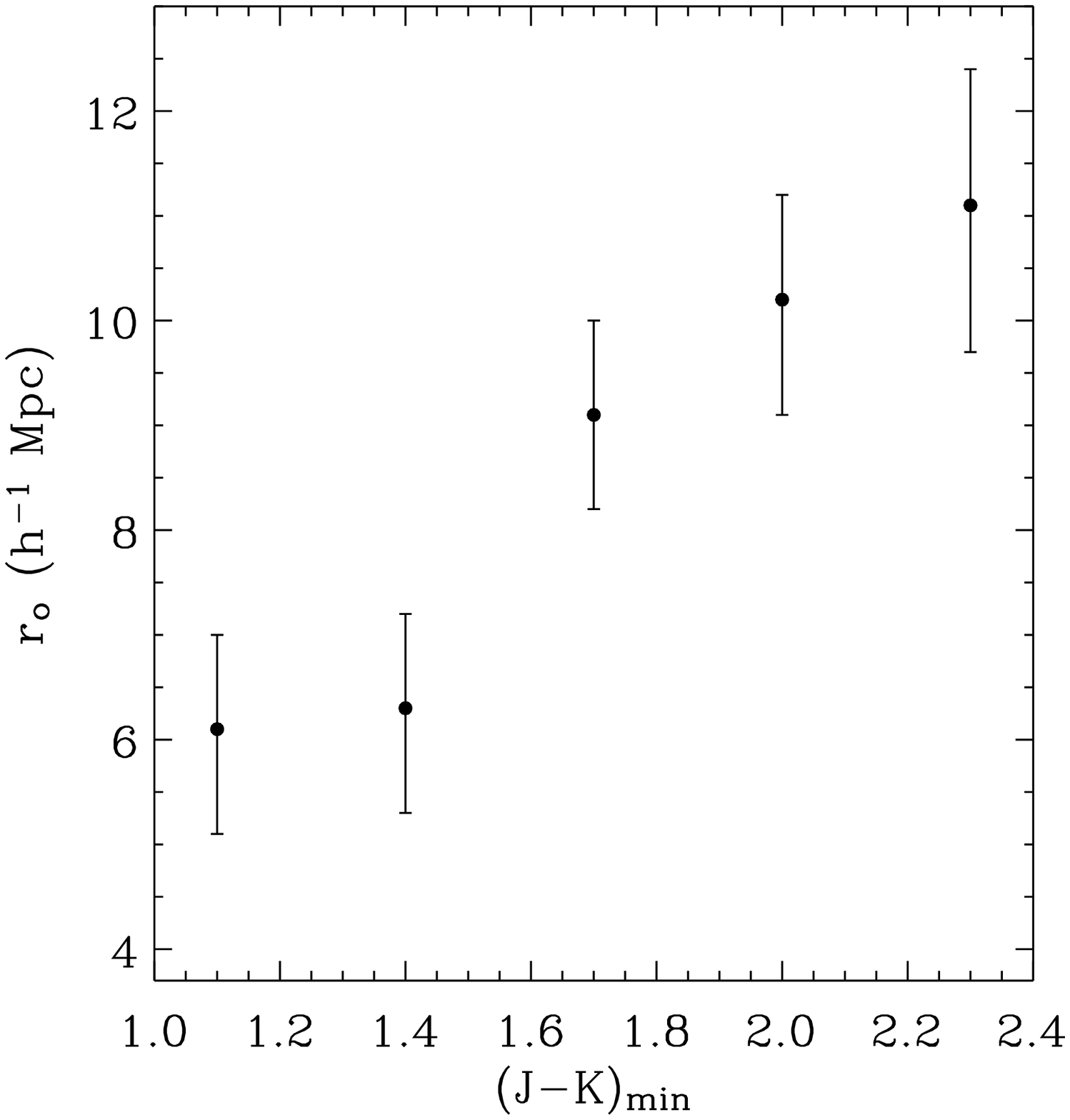}{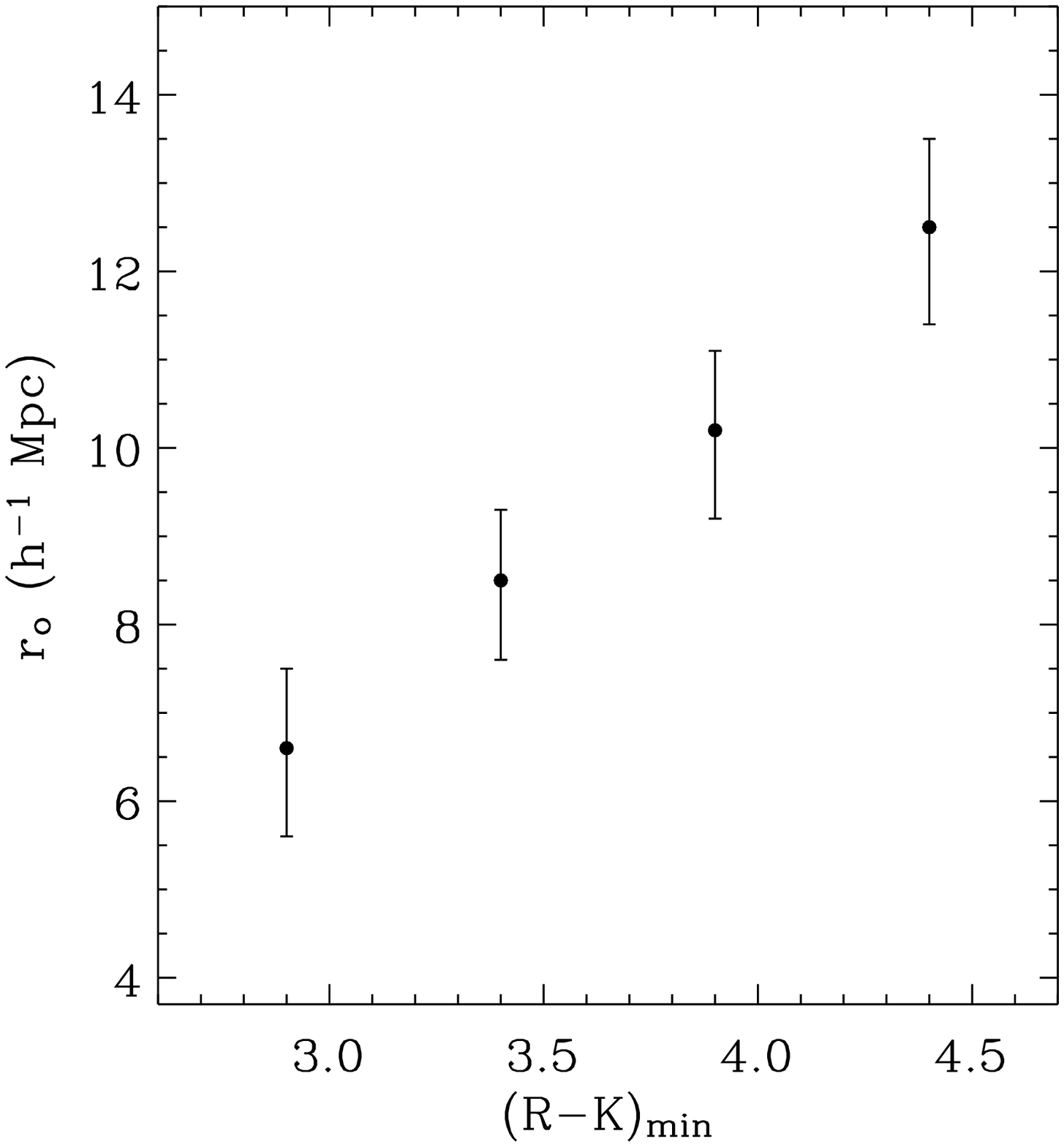}
\caption{The relation between correlation length and color,
for galaxies at $2<z<3.5$ in the MUSYC survey (Quadri et al.\ 2007).
There is a clear correlation, implying that the color-density
relation was already in place at this epoch.
}
\end{figure}

\section{Evidence for a Color-Density Relation at $z\sim 2.5$}

It is not yet possible to study the color-density relation
at $z>2$ with the same tools as has been done at $0<z<1.3$.
Nevertheless, some progress has been made, by employing different
techniques. Steidel et al.\ (2005) compared the masses and
ages of galaxies in an overdensity at $z=2.3$ to those of
other galaxies at the same redshift, with the aid of
Spitzer photometry. Galaxies in the overdensity have
masses and ages which are a factor of $\sim 2$ larger than
those of identically-selected galaxies outside the structure,
implying that galaxies already ``know'' of their large-scale
environment at this early epoch. Continuing on this
theme, Kodama et al.\ (2007) recently
identified a red sequence in overdensities centered on four
radio galaxies at $2<z<3$.

Another approach is to
determine the relative clustering strength of red and blue
galaxies. Daddi et al.\ (2003) found that red galaxies
(selected by their $J-K$ color; Franx et al.\ 2003) in
HDF-South cluster more strongly than blue galaxies.
A similarly high clustering length was derived by Grazian
et al.\ (2006) in the CDF-South. These initial
results were confirmed by Quadri et al.\ (2007), who
for the first time probed sufficiently large areas to study
the clustering of the dark matter halos of red galaxies.
As shown in Fig.\ 5, there is a clear relation between
color and correlation length at $z\sim 2.5$, suggesting that
the color-density relation was already in place at this
redshift.

\section{Summary}

Although environment clearly plays an important role in
galaxy formation and evolution, the similarity of massive
galaxies in clusters and in the general field demonstrates that
other processes are also at work. There is good evidence
that galaxies in clusters and in the field experienced significant
``dry'' merging, but the merger rates need to be determined
better. Studies beyond $z= 1$ offer a somewhat confused
picture, with the color-density relation seemingly absent
at $z\sim 1.3$ and seemingly in place at $z\sim 2.5$.
Ongoing near-IR and Spitzer imaging surveys should greatly
enhance the statistics, and allow direct comparisons of
high redshift galaxies in a wide range of environments. 

\acknowledgements %%% Text of acknowledgements runs on after this command.
The authors wish to thank the organizers for a fun and stimulating
meeting.

%%% THE BIBLIOGRAPHY
%%%
%%% CONSULT SECTION 3 OF "INSTRUCTIONS FOR AUTHORS" FOR HOW TO USE NATBIB.
%%% AUTHORS ARE ENCOURAGED TO USE EITHER THE "THEBIBLIOGRAPY" ENVIRONMENT
%%% BY UNCOMMENTING (DELETING THE "%" SYMBOL) THE COMMANDS BELOW, OR BY
%%% USING THE BIBTEX ENVIRONMENT. TO FIND OUT WHICH IS APPLICABLE TO YOUR
%%% CONTRIBUTION, CONSULT THE VOLUME EDITORS FOR YOUR PROCEEDINGS.
%%%

\end{document}